\documentclass[fleqn,usenatbib]{mnras}
\usepackage{newtxtext,newtxmath}

\usepackage{graphicx}	
\usepackage{amsmath}	
\usepackage{soul}
\usepackage{ragged2e}

\title[SN~2010jp's environment]{Massive stars dying alone:  the remote environment of SN~2010jp and its associated late-time source}

\author[Corgan et al.]{Austin Corgan$^{1}$, Nathan Smith$^{1}$\thanks{E-mail: nathans@as.arizona.edu}, Jennifer Andrews$^{1}$, Alexei V. Filippenko$^{2,3}$, \newauthor Schuyler D.~Van Dyk$^{4}$
\\
$^{1}$Steward Observatory, University of Arizona, 933 North Cherry
  Avenue, Tucson, AZ 85721, USA\\ 
  $^{2}$Department of Astronomy, University of California, Berkeley, CA 94720-3411, USA \\
  $^{3}$Miller Institute for Basic Research in Science, University of California, Berkeley, CA 94720, USA\\
  $^{4}$Caltech/IPAC, Mailcode 100-22, Pasadena, CA 91125, USA
}

\date{Accepted XXX. Received YYY; in original form ZZZ}

\pubyear{2021}

\begin{document}
\label{firstpage}
\pagerange{\pageref{firstpage}--\pageref{lastpage}}
\maketitle

\begin{abstract}
We present late-time images of the site of the peculiar jet-driven Type~IIn supernova (SN)~2010jp, including {\it Hubble Space Telescope} ({\it HST}) images taken 2--5\,yr post explosion and deep ground-based images over a similar time. These are used to characterise its unusually remote environment and to constrain the progenitor's initial mass and age. The position of SN~2010jp is found to reside along a chain of diffuse starlight that is probably an outer spiral arm or tidal tail of the interacting galaxy pair NGC~2207/IC~2163.  There is one bright H~{\sc ii} region projected within 1\,kpc, and there is faint extended H$\alpha$ emission immediately surrounding the continuum source at the position of SN~2010jp, which has $M_{F555W} = -7.7 \pm 0.2$\,mag.  In principle, the lingering light could arise from late-time circumstellar material (CSM) interaction, an evolved supergiant, a host star cluster, or some combination of these. Steady flux over 3\,yr and a lack of strong, spatially unresolved H$\alpha$ emission make ongoing CSM interaction unlikely. If an evolved supergiant dominates, its observed luminosity implies an initial mass $\la 22$\,$M_{\odot}$ and an age $\ga 8$\,Myr.  If the source is a star cluster, then its colour and absolute magnitude imply an age of 8--13\,Myr and a modest cluster initial mass of log($M/M_{\odot}) = 3.6$--3.8.  Extended H$\alpha$ emission out to a radius of $\sim 30$\,pc reveals a faint evolved H~{\sc ii} region, pointing to recent star formation with at least one late O-type star. Based on these various clues, we conclude that the progenitor of SN~2010jp had a likely initial mass of 18--22\,$M_{\odot}$.
\end{abstract}




\section{Introduction}

SN~2010jp was a peculiar Type II supernova (SN) discovered independently in 2010 November by both \citet{maza2010} and as PTF10aaxi by the Palomar Transient Factory 
\citep{rau2009,law2009}. \citet{challis2010} identified it as a Type~IIn SN, and noted very broad line wings outside the narrow emission-line core of H$\alpha$.  
\citet{smith2012} analysed this event in detail and discussed its extreme peculiarity, owing to  its unique triple-peaked H$\alpha$ line profile with a narrow core and extremely broad wings, its relatively low peak luminosity, and its very remote environment.  SNe that arise in remote regions in the outer parts of galaxies or between pairs of galaxies have been of interest for some time \citep{rz67}.  Such isolation is especially mysterious for core-collapse SN subtypes that are thought to arise from unusually massive progenitor stars, whose short lives give them little time to move far from their birth site.  In principle, core-collapse SNe may appear isolated because their modestly massive progenitor formed in relative isolation \citep{oey13,lamb16}, or because the progenitor was a runaway caused by dynamical ejection from a cluster or arising from the kick imparted by a companion star's SN explosion \citep{blaauw61,eldridge2011,gb86,gies87,renzo19,andersson20}; also, the progenitor may appear younger than surrounding stars because it was a blue straggler owing to binary evolutionary effects \citep{bl95,schneider15}.

SNe~IIn are characterised by strong, relatively narrow H$\alpha$ emission lines in their spectra. The narrow core of 100--1000\,km\,s$^{-1}$ is typically accompanied by wings out to $\pm 2000$\,km\,s$^{-1}$, and sometimes broader wings out to $\pm5000$\,km\,s$^{-1}$ or more \citep[e.g.,][]{filippenko1997}. These wings are usually attributed to either electron scattering arising in dense circumstellar material (CSM; especially at earlier times up to and around peak luminosity), or to emission from the expanding post-shock gas or fast SN ejecta at later times \citep{chugai2001,smith2008,smith2020,dessart2015}.  See \citet{smith2017} for a review of SNe with strong CSM interaction.  

The spectra of SN~2010jp exhibited the characteristic narrow line core with a width of 800\,km\,s$^{-1}$, but they also showed broad wings out to more than $\pm 20,000$\,km\,s$^{-1}$ with broad emission peaks centred at $-13,000$ and $+15,000$\,km\,s$^{-1}$, which is extremely unusual.\footnote{One might therefore refer to SN~2010jp as a ``broad-lined Type IIn,'' following the precedent for broad-lined SNe~Ic  (some of which are also associated with jets in gamma-ray bursts).  Since the narrow lines and broad lines arise in physically distinct emitting regions, a broad-lined narrow-lined SN is not as much of a contradiction as it may seem.}  These broad components, added to the narrow core, gave the overall line profile a distinct and unique triple-peaked shape \citep{smith2012}.   Such line profiles may indicate the presence of a collimated jet, making SN~2010jp the first SN~II to be observed with this feature. 

Two proposed mechanisms for the observed triple-peaked H$\alpha$ emission are as follows. (1) $^{56}$Ni is concentrated in collimated polar regions of the H-rich ejecta, perhaps launched there by a fast bipolar jet, where the high-velocity H$\alpha$ arises because H-rich ejecta at the poles are heated by the radioactivity of the $^{56}$Ni. (2) Interactions between a jet and dense CSM result in a prolate reverse shock in the jet which produces red and blue H$\alpha$ emission bumps \citep{smith2012}.  In any case, if a collimated outflow is indeed related to a jet, then the initial mass of SN~2010jp is of great interest.  Some models \citep{heger2003,wh12} predict that at low metallicity, a narrow range of initial masses around 25\,$M_{\odot}$ might produce weak SNe by fallback to a black hole, possibly accompanied by jets if the stars have sufficient rotation and thin H envelopes.  In that context, the possible combination in SN~2010jp of a collimated outflow, a low peak luminosity, and a remote location (perhaps suggesting relatively low metallicity) are tantalising.  Despite its modest peak absolute visual magnitude of only $-15.9$, its high outflow speeds imply a substantial kinetic energy of (1--4) $\times 10^{51}$\,($M_{\rm ej}$/$M_{\odot}$)\,erg.  

Because dense CSM is required, candidates for SN~IIn progenitors include massive stars with high mass-loss rates ($\gtrsim 10^{-2}\,M_{\odot}$\,yr$^{-1}$) characteristic of luminous blue variables \citep[LBVs;][]{kiewe2012,smith2014,smith2017,smith2017b}.  Though very massive stars have short main-sequence lifetimes and may be expected to reside near young star-forming regions, SN~2010jp appears to be quite isolated.  Some other SNe~IIn, such as SN~2009ip, have been observed in similarly remote environments \citep{smith2016}, even though SN~2009ip had a very luminous progenitor that seemed consistent with a 50--80\,$M_{\odot}$ star \citep{smith2010}. Luminous, massive stars that are isolated from other massive stars or appear older than surrounding stars are difficult to reconcile with expectations of single-star evolution.  Anomalously isolated massive stars may arise, however, through binary interaction where mass accretion or mergers produce blue stragglers, or where a star may receive a kick from a companion's SN, as noted above.  Like some SNe~IIn, LBVs themselves are also observed to be unexpectedly isolated compared to massive stars of comparable luminosity, probably as products of binary evolution \citep{smith2015,mojgan2017}.  Other potential mechanisms for strong pre-SN mass loss may not require the highest initial masses, including wave driving or other instabilities associated with very late nuclear burning phases \citep{qs2012,smith2014,sa14,fuller2017,fr18,lf20} or violent binary interaction shortly before explosion \citep{sa14}.  The progenitor star's initial mass may have important implications for the unusual explosion properties of SN~2010jp, as noted above, and so here we take a close look at the remote environment of SN~2010jp long after the SN explosion has faded.

We adopt a distance to SN~2010jp of 24.5\,Mpc ($m-M = 32.9$\,mag; \citealt{smith2012}), where it resides at coordinates $\alpha_{\rm J2000} = 06^{\mathrm{h}}16^{\mathrm{m}}30\fs63$, 
$\delta_{\rm J2000} = -21^\circ24'36\farcs2$. It is in a remote environment more than 30\,kpc from the centres of the merging pair of star-forming galaxies IC~2163 and NGC~2207 (Fig.~\ref{fig:image}).  See \citet{smith2012} for observational details of the SN event.   In this paper, we further characterise SN~2010jp at late times and its remote environment by constraining the age and initial mass via analysis of the remaining source and its surroundings.   In Section 2 we present late-time ground-based and {\it Hubble Space Telescope (HST)} images, and in Section 3 we present our main results, including photometry and extended 
structure analysis of the remaining source at the location of SN 2010jp. Section 4 discusses  candidates for an associated late-time source and gives likely mass and age estimates for that source. We conclude in Section 5.

\section{Observations}

\subsection{Ground-based images}

\begin{figure*}
    \includegraphics[width=\textwidth]{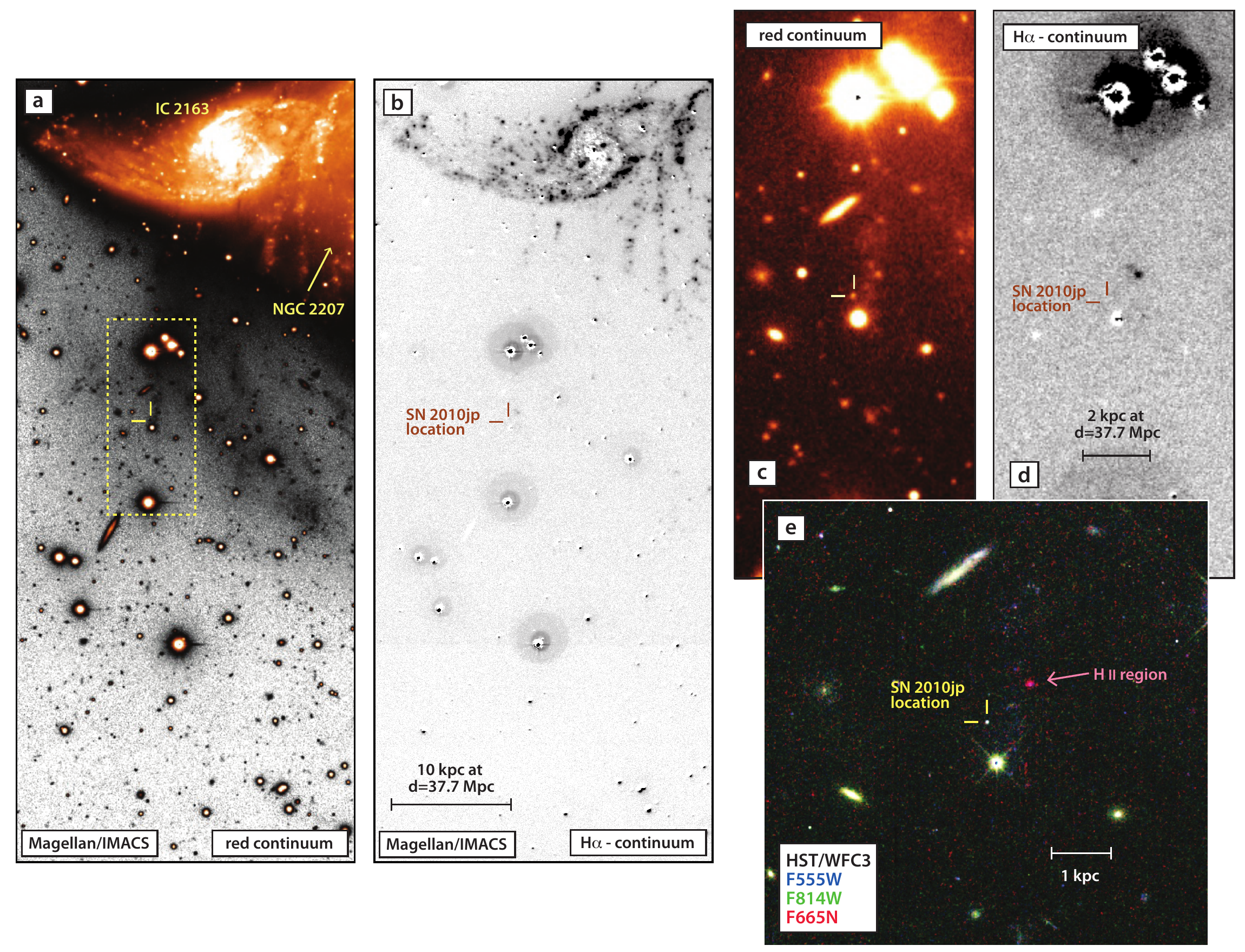}
    \caption{ Panels (a) and (b) show relatively wide-field images of the environment of SN~2010jp taken with IMACS on Magellan in 2015, centred south of the merging pair of galaxies IC~2163 (at top) and NGC~2207 (nucleus out of this field to the upper right).  Panel (a) is a wide-band red continuum image with bright features in a positive orange intensity scale, and with the faintest features in a negative greyscale to emphasise outer tidal debris of the merging galaxies. Panel (b) shows the same field but with continuum-subtracted H$\alpha$ emission in negative greyscale, made by subtracting the image in Panel (a) from a narrow-band 6600\,\AA\ image that includes H$\alpha$ at the redshift of SN~2010jp (see text).  Panels (c) and (d) are similar to (a) and (b), but zoomed-in on a smaller field of the same Magellan/IMACS images in the vicinity of SN~2010jp, marked by the dashed rectangle in Panel (a), and with a simple positive orange-tinted intensity scale in (c). Panel (e) shows a 3-colour image made from {\it HST}/WFC3 images obtained in 2015, with {\it F555W} in blue, {\it F814W} in green, and H$\alpha$ ({\it F665N}) in red.  An H~{\sc ii} region located about 1~kpc northwest of the location of SN~2010jp is seen as a red feature in Panel (e) and as a subtraction residual in Panel (d).  In each panel, crosshairs mark the location of SN~2010jp and/or its associated late-time source.
}
    \label{fig:image}
\end{figure*}

We obtained late-time wide-field images of the environment of SN 2010jp in both red continuum and narrow H$\alpha$ filters with Magellan/IMACS \citep{2011PASP..123..288D} on three separate nights:  2013 April 6, 2014 February 6, and 2015 January 20 (UT dates are used throughout this paper). The red continuum filter was a wide bandpass (WBP) filter spanning 6226--7171\,\AA, and the narrow-band filter was MMT6600/260 (with central $\lambda=6600$\,\AA\, and width $\Delta\lambda=260$\,\AA).  This narrow-band filter included H$\alpha$ at the SN redshift of $z=0.009$ \citep{smith2012}. At each of the three epochs, we obtained a series of exposures totalling a minimum of 720\,s in the WBP red filter and 1800\,s in the narrow H$\alpha$ filter.  The seeing was subarcsecond and conditions were photometric on all of the nights.  Standard reduction procedures in {\tt IRAF}\footnote{{\tt IRAF}, the Image Reduction and Analysis Facility is distributed by the National Optical Astronomy Observatory, which is operated by the Association of Universities for Research in Astronomy
(AURA), Inc., under cooperative agreement with the U.S. National Science Foundation (NSF).} were used for bias correction and flatfielding.  A faint point source was detected in all images consistent with the location of SN~2010jp, and photometry of this source is discussed below in Section 3.3.  Figure~\ref{fig:image} shows the WBP red continuum and continuum-subtracted H$\alpha$ images from 2015. The environment around SN~2010jp is discussed more below.

\subsection{Late-time \textit{HST} Images}

\textit{HST} observations of SN2010jp were taken with WFC3/UVIS on 2012 December 4 using the {\it F814W} filter as part of SNAP proposal GO-13029 (PI A. V. Filippenko). The data were obtained from the Mikulski Archive for Space Telescopes (MAST\footnote{https://archive.stsci.edu/}) with standard pipeline calibrations applied.  We observed SN~2010jp again with {\it HST} using WFC3/UVIS on 2015 May 14 with filters {\it F555W}, {\it F665N}, and {\it F814W} (GO-13787, PI N. Smith). The {\it F665N} filter was chosen to include H$\alpha$ at the SN redshift of $z=0.009$ \citep{smith2012}.  Exposure times and other details are listed in Table~\ref{tab:photparams}.

A single point source was clearly detected in all images consistent with the position of SN~2010jp. Photometry was performed on the FLC frames (*.flc files are UVIS-calibrated and CTE-corrected frames) using DOLPHOT\footnote{http://americano.dolphinsim.com/dolphot/} \citep{dolphin2000,2016ascl.soft08013D}, from which the object type was indicated to be stellar.  For the DOLPHOT parameter file we have used the  recommended values, including FitSky=2, RAper=3, and RPSF=13.  The resulting \textit{HST} photometry is listed in Table \ref{tab:photparams}.

\begin{table*}
	\centering
	\caption{Photometry parameters and results for late-time \textit{HST} images of SN 2010jp.}
	\label{tab:photparams}
	\begin{tabular}{lcccc} 
		\hline\hline 
		 & $F555W$ & $F665N$ & $F814W$ & $F814W$ (2012)\\
		\hline
		Date (MJD) & 57156.26 & 57156.13 & 57156.34 & 56265.52\\
		Exposure time (s) & 4219 & 5892 & 1528 & 680\\
		$\lambda_{\mathrm{eff}}$ (\AA) & 5234.6 & 6656.4 & 7977.1 & 7977.1\\
		$f_{\lambda} \left( 10^{-19}\mathrm{\frac{erg}{cm^2\,s\,\text{\AA}}} \right)$ & 5.93 (0.08) & 6.1 (0.2) & 3.8 (0.1) & 4.0 (0.1)\\
		Vega magnitude & 24.56  (0.01) & 23.77 (0.04) & 23.70 (0.03) & 23.64 (0.04)\\
		\hline
\end{tabular}

{\justify Note: Images were taken in 2015 with the exception of the right column (taken in 2012). Photometry was performed with DOLPHOT using recommended values (FitSky=2, RAper=3, RPSF=13; as a check, we also measured photometry using RAper=8 and FitSky values of 1 and 3, which gave fluxes 1$\sigma$ and 2$\sigma$ larger, respectively). The $f_{\lambda}$ and Vega magnitude values as reported in this table are prior to any reddening corrections. Of particular interest is that there is no significant change in the flux measured in the {\it F814W} filter from 2012 to 2015, indicating that the observed source is not likely dominated by ongoing late-time CSM interaction from the SN event. \par}
\end{table*}

\section{Results}

\subsection{Environment around SN~2010jp}

The images of SN~2010jp provide some basic qualitative information about its local environment.  The ground-based continuum image (Fig.~\ref{fig:image}a and \ref{fig:image}c) shows low surface brightness emission from outer spiral arms of NGC~2207 or tidal tails from the interacting pair of galaxies.  SN~2010jp appears to reside close to (within 1 kpc) of one such faint string of continuum emission running north to south.  The {\it HST} images in {\it F555W} and {\it F814W} indicate that the stars seen in this tidal tail have a relatively blue colour, perhaps indicating a moderately young population.  This suggests that SN~2010jp belongs to one of these outer tails of the merging spirals, rather than to a very faint satellite dwarf galaxy.

While the continuum-subtracted H$\alpha$ image (Fig.~\ref{fig:image}b and \ref{fig:image}d) shows copious star formation via chains of H~{\sc ii} regions along spiral arms in IC~2163 and NGC~2207, there is little evidence of bright H$\alpha$ emission in the far outer spiral arms more than 10\,kpc from the galaxy centres.  There is, however, one moderately bright H$\alpha$ source located about 1\,kpc to the northwest of SN~2010jp (Fig.~\ref{fig:image}d), which is also seen as a red source in the {\it HST} colour images (Fig.~\ref{fig:image}e) where {\it F665N} (H$\alpha$) is coloured red. While we do not see obvious H$\alpha$ emission at the position of SN~2010jp in the continuum-subtracted or colour images in Fig.~\ref{fig:image}, photometry does reveal a slight excess of emission in the {\it F665N} filter (see below, Sec. 3.3), and an analysis of the point-spread function (PSF) indicates that the {\it F665N} emission is extended (see below, Sec. 3.2).  It is therefore likely that SN~2010jp itself resides in a faint H~{\sc ii} region.  This, in turn, means that the relative isolation of SN~2010jp most likely does not arise because it has travelled far from its birth site.

Given the general lack of evidence for widespread ongoing star formation, one might naturally expect relatively low local extinction in these remote outer parts of the host galaxies.  Background spiral galaxies are seen near SN~2010jp in the ground-based continuum image and in {\it HST} images, also suggesting rather low local extinction in these outer parts of IC~2163  and NGC~2207.

\subsection{Extended structure}

\begin{figure}
    \centering
	\includegraphics[width=2.8in]{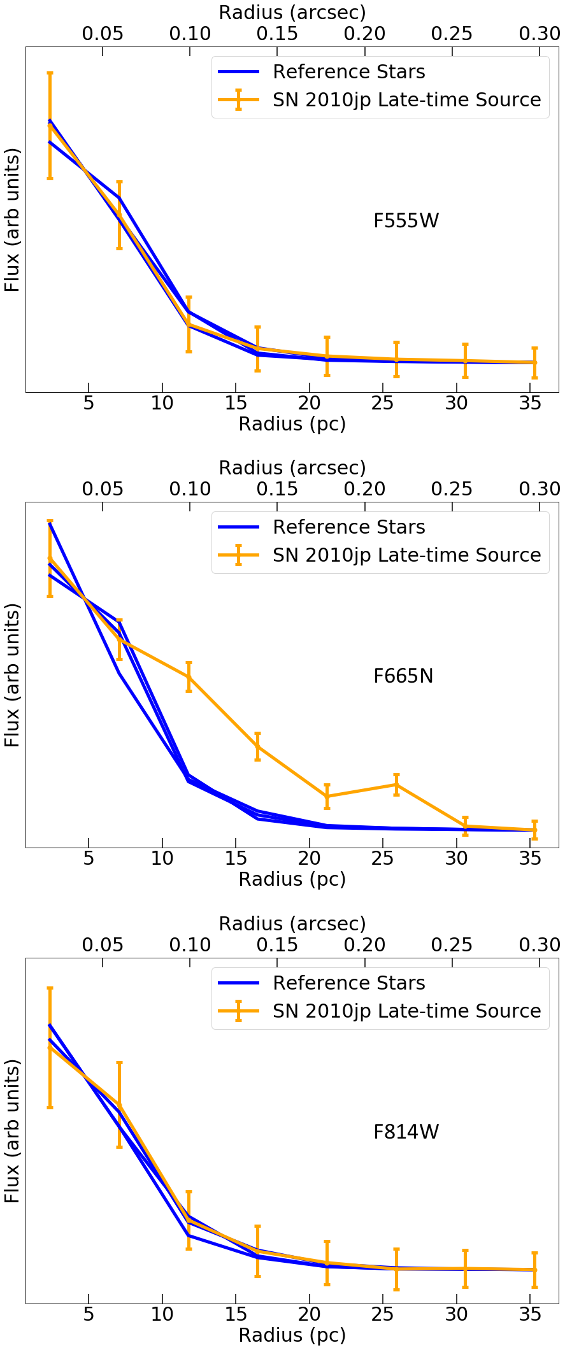}
    \caption{Radial profiles of the late-time source at the location of SN 2010jp (orange) along with PSFs of three reference stars in the field of view (blue) for the {\it F555W}, {\it F665N}, and {\it F814W} filters obtained from 2015 {\it HST} images. PSFs are scaled such that the enclosed flux out to $\sim 10$\,pc is the same for all PSFs in a given filter. In the continuum filters ({\it F555W} and {\it F814W}), PSFs for the source are consistent with those of reference stars, but show significant signal above that of reference stars in the H$\alpha$ filter {\it F665N}.  This is indicative of an extended H~{\sc ii} region around the SN's location out to $\sim$30\,pc.}
    \label{fig:psfs}
\end{figure}

We would like to know whether the point-like source at the position of SN~2010jp is consistent with a true unresolved point source or if there is some extended structure.  A width larger than a PSF in broad-band continuum filters might indicate a dispersing cluster or loose association, whereas extended emission in the {\it F665N} filter might correspond to extended H$\alpha$ emission from an H~{\sc ii} region.   Radial profiles of the intensity of the late-time source at the SN's location are shown in Figure \ref{fig:psfs} for each filter of the 2015 images, compared with PSFs derived from three reference stars in the nearby field of view included in the same images. The PSFs are normalised such that the area under the curve out to $\sim 10$\,pc is the same for each PSF in a given filter. The uncertainty is calculated as $\sigma/\sqrt{n}$, where $\sigma$ is the standard deviation in counts given by the IRAF \textit{phot} command and $n$ is the number of pixels in the radial annulus. 

Radial profiles of the remaining source at the position of SN~2010jp as seen in the {\it F555W} and {\it F814W} filters are consistent with reference-star PSFs, indicating that the continuum luminosity arises from a source that is unresolved and less than $\sim$10 \,pc across. However, there is clear evidence of extended H$\alpha$ emission out to a radius of $\sim$30\,pc in the {\it F665N} filter. This indicates that the SN's location is embedded within extended H$\alpha$ emission, probably coming from a relatively small and faint H~{\sc ii} region.  This H~{\sc ii} region's emission is evidently too faint to be detected in our ground-based images (Fig. \ref{fig:image}), perhaps because this ground-based  MMT6600/260 filter is rather wide and includes about twice as much continuum as the $F665N$ filter. This extended H$\alpha$ emission implies that the position of SN~2010jp is coincident with recent star formation, and again signifies that a significant kick is therefore not the reason that SN~2010jp appears isolated.  This is discussed further below.

\subsection{Photometry}

A faint point source consistent with the position of SN~2010jp was detected in all {\it HST} and ground-based IMACS images (see Fig.~\ref{fig:image}). In Table \ref{tab:photparams}, we provide $f_{\lambda}$ and Vega magnitude values derived from PSF-fitting photometry for the late-time source in all {\it HST} images. It should be noted that $f_{\lambda}$ did not vary significantly in the {\it F814W} filter from 2012 to 2015, although we would expect some significant fading if a late-time SN source dominates the luminosity 2--5\,yr after explosion. Similarly, in Table~\ref{tab:imacsmags} we provide Vega magnitudes for the point source as seen in ground-based IMACS images, also measured with PSF-fitting photometry.  As in the {\it HST} photometry, the magnitude of the observed source did not vary significantly between 2013 and 2015 in the ground-based Magellan/IMACS images (Table \ref{tab:imacsmags}).

\begin{figure}
    \includegraphics[width=\columnwidth]{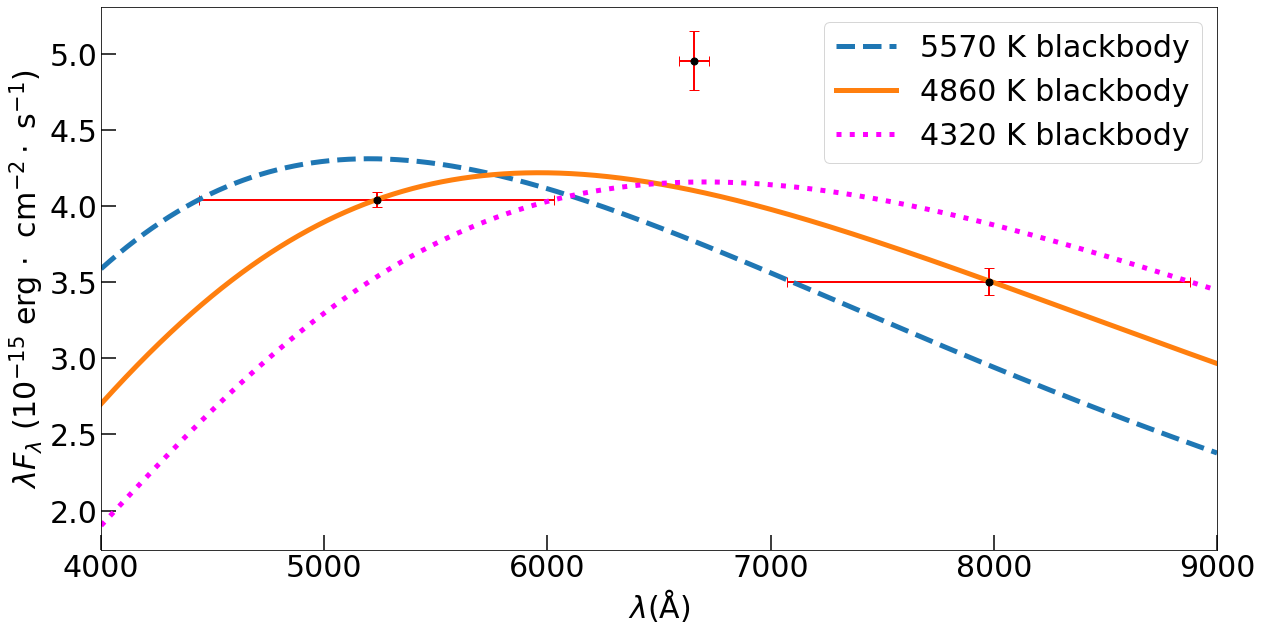}
    \caption{SED of the late-time source at the location of SN~2010jp obtained using 2015 {\it HST} images. Data points have been corrected for Milky Way reddening. Horizontal error bars indicate effective filter widths. Scaled blackbody distributions for three temperatures are included, and 4860$^{+710}_{-540}$\,K is taken as the characteristic temperature of the source for analysis. }
    \label{fig:sed}
\end{figure}

Values of $\lambda f_{\lambda}$ from the {\it HST} images, corrected for the adopted value of Milky Way reddening, are shown in a spectral energy distribution (SED) in Figure~\ref{fig:sed}. The flux in the H$\alpha$ filter can be seen to be $\sim 20$\% above the approximate continuum level as established by representative blackbody curves passing through the broad filter measurements. Milky Way reddening correction was accomplished by adopting $A_V = 0.27$\,mag and $E(B-V) = 0.087$\,mag from  \cite{schlegel1998}\footnote{https://irsa.ipac.caltech.edu/applications/DUST/} and using the algorithm by \cite{cardelli1989}\footnote{http://www.dougwelch.org/Acurve.html} to compute reddening factors at the $\mathrm{\lambda_{eff}}$ associated with each filter.

\begin{table}
    \centering
    \caption{Vega mags of the source detected at the location of SN~2010jp.} 
    \begin{tabular}{lcc}
        \hline\hline 
        Date & NB mag & WB mag \\
        \hline 
        2013 Apr 6  & 23.10 (0.13) & 23.37 (0.18) \\
        2014 Feb 6  & 23.16 (0.15) & 23.01 (0.14) \\
        2015 Jan 20 & 23.09 (0.14) & 23.60 (0.16) \\
        \hline
    \end{tabular}
    \label{tab:imacsmags}
    
    {\justify Note: Uncertainties are 1$\sigma$. Data obtained with Magellan/IMACS between 2013 and 2015, measured with PSF-fitting photometry. The filters are the narrow-band (NB) MMT6600/260 (central $\lambda = 6600$\,\AA, width $\Delta\lambda = 260$\,\AA) and a red continuum wideband (WB) filter (6626--7171\,\AA). The lack of substantial fading is consistent with the \textit{HST} images (Table \ref{tab:photparams}), again indicating that the source is likely dominated by something other than late-time CSM interaction from the SN event. \par}
\end{table}

\subsection{Implied luminosity and temperature} \label{subsec:lumandtemp}

Using $E(B-V)= 0.087$\,mag to correct for Milky Way reddening, we find the absolute {\it F555W} magnitude to be ${M_{F555W}} = -7.7 \pm 0.2$\,mag, with the error bar obtained by assuming a $10\%$ distance uncertainty. Taking this as the absolute bolometric magnitude (i.e., making no additional bolometric correction) yields $\log(L/L_{\odot}) = 4.96 \pm 0.09$ for the late-time source associated with SN~2010jp. Of course, there could be additional reddening and extinction from local dust within the host galaxy or circumstellar dust, which would raise the luminosity.  The colour of the SN \citep{smith2012} and the remote environment suggest that any local extinction is modest.  Adopting an additional hypothetical reddening correction of $E(B-V) = 0.1$\,mag ($A_V = 0.31$\,mag) to account for some possible local reddening at the location of SN~2010jp would yield a luminosity as high as $\log(L/L_{\odot}) = 5.08 \pm 0.09$.

Scaled blackbody distributions are plotted along with the SED in Figure \ref{fig:sed}. We adopt 4860$^{+710}_{-540}$\,K as the approximate continuum temperature of the source.  If we assume the hypothetical additional local reddening correction of $E(B-V) = 0.1$\,mag mentioned above, a similar analysis would yield a slightly higher temperature of 5220$^{+720}_{-550}$\,K.

\section{Discussion}

Three possibilities for the nature of the lingering late-time source at the location of SN 2010jp are plausible: (1) ongoing late-time CSM interaction from the SN event; (2) an individual evolved supergiant star that dominates the visual-wavelength flux; or (3) a host cluster or association.  Although late-time CSM interaction may contribute to the observed source to some extent, we discount option (1) as the dominant source of luminosity for two reasons.  First, if CSM interaction dominated the observed flux, the source would be expected to fade significantly between the observations made in 2012 and 2015, and this is not the case (Table \ref{tab:photparams}). Second, such late-time CSM interaction is unlikely to produce a significant continuum luminosity and would instead produce strong, broad H$\alpha$ emission.  While we do detect some H$\alpha$ emission in excess of the continuum, this emission is resolved and extends out to $\sim$30\,pc from the source (Fig.~\ref{fig:psfs}) and likely arises from a surrounding H~{\sc ii} region.  H$\alpha$ emission from late interaction of SN~2010jp would be unresolved at an age of only a few years after explosion.  The more likely options (2) and (3) are considered in greater detail below.

\subsection{An individual evolved star} \label{sec:individualstar}

\begin{figure}
    \includegraphics[width=\columnwidth]{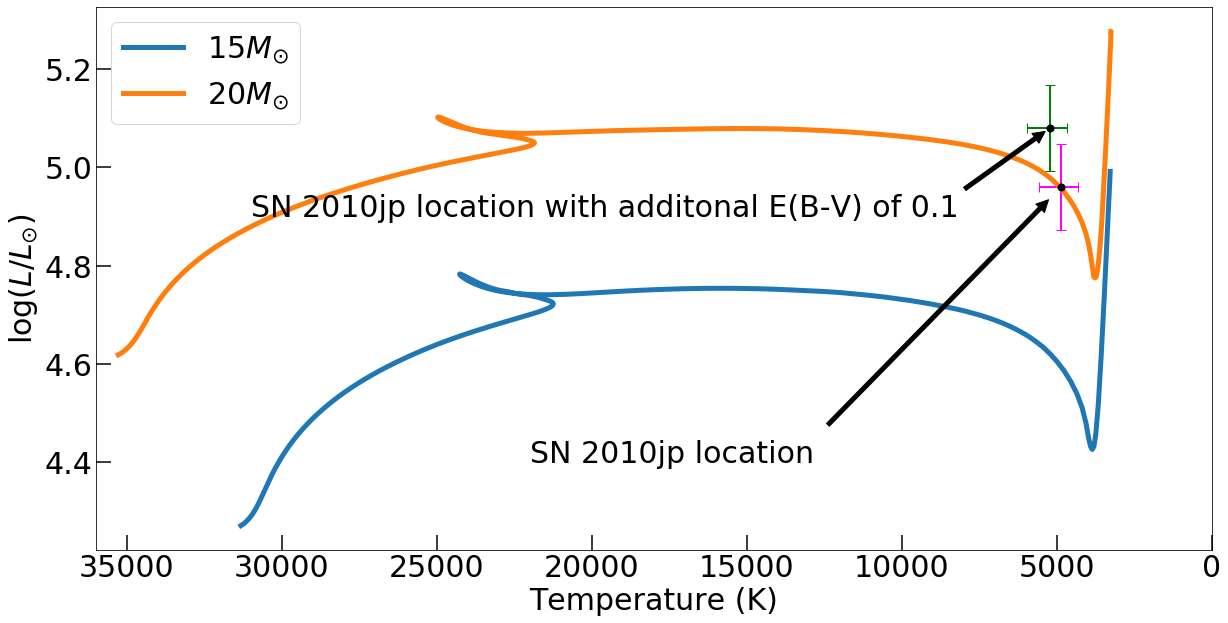}
    \caption{Placement of the late-time source at the location of SN~2010jp on an HR diagram with stellar-evolution tracks for 15 and 20\,$M_{\odot}$ stars. One data point is corrected only for Milky Way reddening, and a more luminous point is shown for which an additional $E(B-V) = 0.1$\,mag is assumed to account for local reddening. Error bars are the result of assuming a 10$\%$ distance uncertainty. Comparison with the stellar-evolution tracks indicates that the temperature and luminosity of the late-time source at the SN's location are consistent with a single evolved star of mass 15--20\,$M_{\odot}$.}
    \label{fig:hrd}
\end{figure}

Assuming that SN 2010jp's associated late-time source is a single evolved star allows its placement on a Hertzsprung-Russell (HR) diagram. Figure \ref{fig:hrd} plots the luminosities and temperatures determined in \ref{subsec:lumandtemp} along with stellar-evolution tracks for 15 and 20\,$M_{\odot}$ stars \citep{brott2011}. In the case that the late-time source is dominated by a single evolved star, we estimate that this star would be of mass 15--20\,$M_{\odot}$ by inspection of Figure \ref{fig:hrd}. The luminosity corresponds to the endpoint of evolution for a 15\,$M_{\odot}$ star (or slightly higher for some additional local extinction), or to a range of late evolutionary times for a star up to 20\,$M_{\odot}$.  This mass estimate will be used to restrict the age of the source in Section \ref{sec:finalestimates}.  In particular, the assumption that all the light comes from one supergiant that evolved as a single star provides an upper limit to the initial mass of that star, and hence a lower limit to the age.  If that star evolved through binary interaction or if the light is a combination of stars, then the true age could of course be older.

\subsection{Extended emission} \label{sec:extendedemission}

In continuum filters ({\it F555W} and {\it F814W}), the radial profiles of the source at the position of SN~2010jp are consistent with PSFs determined from reference stars, while in the H$\alpha$ filter ({\it F665N}) the source is significantly extended, with clear excess flux above reference-star PSFs at projected radii out to $\sim 0.25''$ or $\sim 30$\,pc (Fig. \ref{fig:psfs}). This is consistent with the fact that H$\alpha$ emission is observed to be above the continuum level (Fig. \ref{fig:sed}), and is most likely indicative of a surrounding H~{\sc ii} region. The presence of an extended H~{\sc ii} region strengthens the hypothesis that a host star cluster dominates the light from the late-time source at the SN's location. It also rules out the hypothesis that SN~2010jp appears isolated because it has moved far from its birthplace, as noted earlier. 

An H~{\sc ii} region radius of only $\sim 30$\,pc is quite modest compared to the $\sim 100$\,pc giant H~{\sc ii} regions and superbubbles that surround very massive stars and young massive clusters.  This may point to a relatively modest initial mass for the progenitor of SN~2010jp, also suggesting that it was not a lower-mass sibling of much more massive stars that had already exploded.  The observed radius of $\sim 30$\,pc can be compared to the expected Str\"omgren sphere radius for an O-type star.  The size of a Str\"omgren sphere can be expressed as

\begin{displaymath}
R_{\rm pc} \approx 31 \, \Big{(} \frac{Q}{10^{48}} \Big{)}^{\frac{1}{3}} \, n_e^{-\frac{2}{3}},
\end{displaymath}

\noindent where $Q$ is the ionising photon rate in photons\,s$^{-1}$ produced by an O-type star and $n_e$ is the ambient electron density (cm$^{-3}$) \citep{stromgren1939}. The ambient density is not expected to be unusually high in such a remote outer region of a galaxy, so taking a representative density of $\sim 1$\,cm$^{-3}$, the observed H~{\sc ii} region radius would require an ionising source with $Q \approx 10^{48}$\,photons\,s$^{-1}$ for most of its main-sequence life.  This corresponds to a late O-type star (roughly O8 to O9 V; \citealt{vacca96,martins02}), which corresponds to an initial mass of roughly 18--20\,$M_{\odot}$.   We note that a handful of late O-type stars in the Small Magellanic Cloud (SMC) that appear isolated are observed to have circular H~{\sc ii} regions with similar physical radii \citep{oey13}.   Although this is admittedly a quite imprecise constraint on the initial mass of the SN's progenitor because we have simply adopted a representative ambient density, it is nevertheless consistent with the initial mass range inferred from other diagnostics in this work. 

The presence of a surrounding H~{\sc ii} region therefore points to a moderately high mass of at least 18-20\,$M_{\odot}$, even though the H$\alpha$ flux is quite faint compared to H$\alpha$ emission from the inner parts of the host galaxies. The H$\alpha$ emission from this H~{\sc ii} region is only detected in deep images with {\it HST}, but is not detected in deep continuum-subtraction images obtained with a ground-based 6.5\,m telescope.  This suggests that the faintness of the H$\alpha$ from this environment is a consequence of the low interstellar medium (ISM) densities and overall low star formation rates found in the 
remote outer parts of the host galaxies, not because of a relatively low initial mass around 8\,$M_{\odot}$.  This, in turn, suggests that metrics which equate the brightness of local H$\alpha$ emission with youth or use ordered flux ranking of H$\alpha$ pixels to infer relative ages \citep{aj08} may not be good indicators of the SN progenitor star's initial mass, and may instead be influenced by location in a galaxy as well as metallicity.

\subsection{Host cluster age}
\label{sec:cluster}

\begin{figure*}
	\includegraphics[width=6.0in]{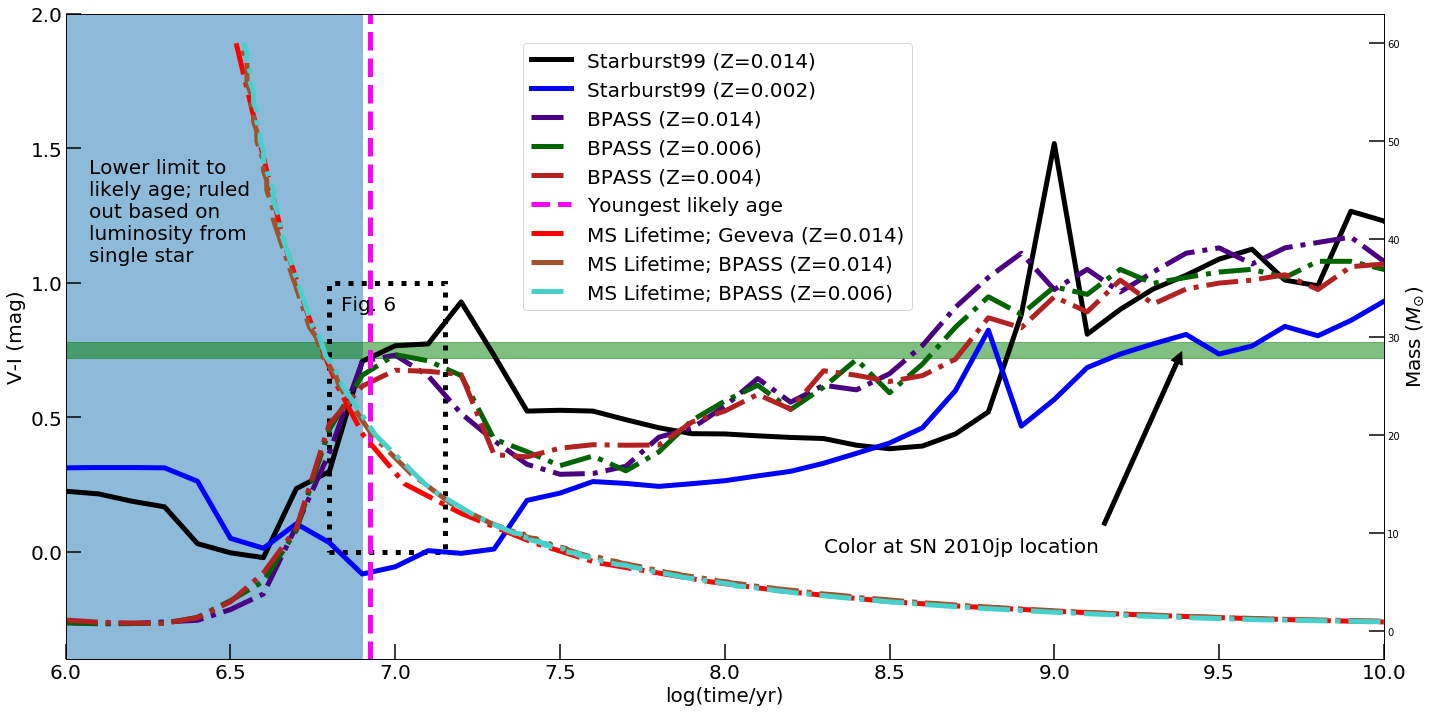}
    \caption{The observed $V-I$ colour (corrected for Milky Way reddening) at SN~2010jp's location is shown as the shaded green horizontal bar. Also shown are simulated $V-I$ colours as a function of time from Starburst99 and BPASS cluster evolution models (single-age; $10^6\,M_{\odot}$ total mass; Salpeter IMF 0.1--100\,$M_{\odot}$) with the indicated metallicities. We also show curves based on Geneva and BPASS stellar-evolution models that indicate stellar masses (right axis) leaving the main sequence as a function of time. We estimate the likely age of a cluster consistent with the observed late-time source at SN~2010jp's location based on where the observed colour is consistent with the models. A single star more massive than $20\,M_{\odot}$ would give a luminosity higher than that observed, and would leave the main sequence at $\sim 8$\,Myr.  Thus, this is assumed to be a lower limit on the age; the excluded region is shaded blue on the left side of the figure. Starburst99 ($Z=0.014$) becomes consistent with the observed colour at $\sim 8.4$\,Myr, making this the youngest likely age for a cluster consistent with the observed late-time source associated with SN~2010jp.}
    \label{fig:main}
\end{figure*}

Taking the {\it F555W} and {\it F814W} magnitudes from 2015 (Table \ref{tab:photparams}) as $V$-band and $I$-band magnitudes (respectively) and adjusting for Milky Way reddening as in \ref{subsec:lumandtemp}, we find a $V-I$ colour of $0.75 \pm 0.03$\,mag; this colour is indicated by the horizontal green bar in Figure \ref{fig:main}. Also included are the colours given by cluster-evolution models from Starburst99\footnote{https://www.stsci.edu/science/starburst99/docs/default.htm} \citep{leitherer1999,leitherer2005,leitherer2010,leitherer2014} and BPASS\footnote{https://bpass.auckland.ac.nz/} \citep{stanway2018,eldridge2017}. Both Starburst99 and BPASS simulate single-aged stellar populations of total mass $10^6$\,$M_{\odot}$ with Salpeter initial mass functions \citep[IMFs;][]{salpeter1955} for individual stellar masses 0.1--100\,$M_{\odot}$. The Starburst99 models use zero-rotation Geneva (2012/2013) stellar-evolution tracks; BPASS uses its own custom tracks. Colours are shown for various metallicities: $Z=0.014$, 0.006, and 0.004 for BPASS, and $Z=0.014$ and 0.002 for Starburst99 (these are the only two metallicities available with the Geneva tracks used). Also shown are initial stellar masses (right axis) as a function of main-sequence lifetime for $Z=0.014$ Geneva (i.e., Starburst99) models, as well as for $Z=0.014$ and 0.006 BPASS models. These times are estimated for the Geneva models as suggested by \cite{ekstrom2012}, and taken directly from the BPASS website\footnote{https://bpass.auckland.ac.nz/7.html} for those models. All estimates of main-sequence lifetimes in this section and the next are obtained similarly, referencing the appropriate lifetimes for the model under consideration. We now compare the observed colour with these models in order to estimate the age and mass of a host cluster consistent with our observations, as described below.

\begin{figure*}
    \includegraphics[width=6.0in]{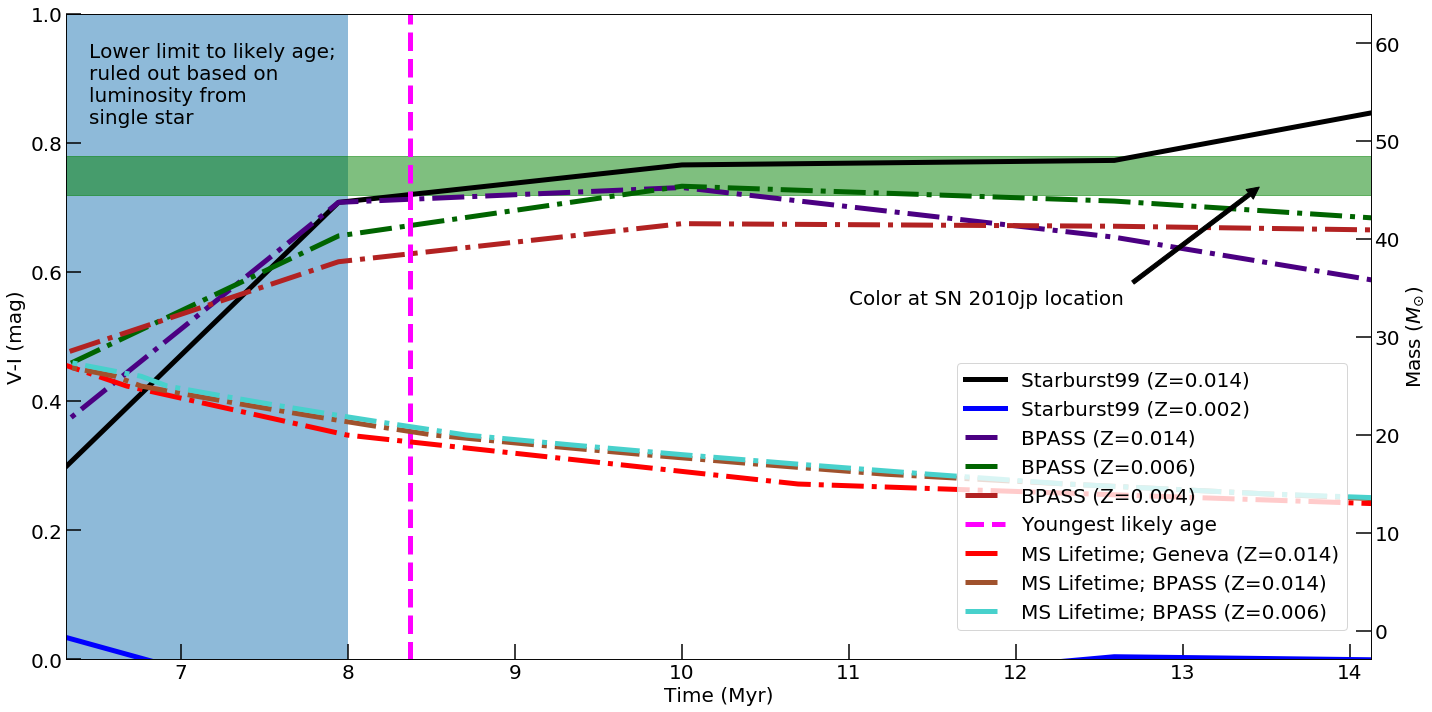}
    \caption{Expansion of the crucial region of Figure \ref{fig:main} with a linear time scale. The simulated $V-I$ colour from Starburst99 ($Z = 0.014$) is consistent with observations from 8.4 to 12.7\,Myr, BPASS ($Z=0.014$) is consistent from 9.0 to 10.4\,Myr, and BPASS ($Z=0.006$) is consistent from 9.7 to 11.5\,Myr.}
    \label{fig:mainzoom}
\end{figure*}

It can be seen in Figures \ref{fig:main} and \ref{fig:mainzoom} that Starburst99 ($Z=0.014$) is consistent with the observed colour from 8.4 to 12.7\,Myr. BPASS ($Z=0.014$) is consistent from 9.0 to 10.4\,Myr, and BPASS ($Z=0.006$) from 9.7 to 11.5\,Myr. The next time any model is consistent with the observed colour is Starburst99 ($Z=0.014$) at $\sim 19$\,Myr, but this corresponds to the main-sequence lifetime of a $\sim 11$\,$M_{\odot}$ star according to both Geneva and BPASS models. All O-type stars would be long gone by this point. However, late O-type stars are required to ionise the extended H~{\sc ii} region that is, indeed, observed (Fig. \ref{fig:psfs}). Overall, we take 8--13\,Myr as our estimate for the age of a host cluster. We note that the BPASS ($Z=0.004$) and Starburst99 ($Z=0.002$) models do not become consistent with observations until times far too late to be reasonable estimates for the cluster age ($>300$\,Myr). This indicates that, while the cluster metallicity may indeed be mildly subsolar as expected, it is likely not at or below SMC-like metallicity ($\lesssim 0.004$). We will primarily focus on the $Z=0.014$ Starburst99 and $Z=0.014$ and 0.006 BPASS models in our discussion. 

Following \cite{werk2008}, we can obtain an estimate for the initial mass of the cluster by comparing the observed $V$-band magnitude $M_V$ with the $V$ magnitudes predicted by the Starburst99 and BPASS models $M_{V,{\rm model}}$ at the estimated age, and then determining the factor by which the mass would need to be scaled to obtain our observed $M_V$. Noting that both models assume an initial cluster mass of $10^6\,M_{\odot}$, we can obtain an estimate for the cluster mass from $10^{6+0.4(M_{V,{\rm model}}-M_V)}$. After correcting for Milky Way reddening, we obtain an observed $M_V \approx -7.7$\,mag. Throughout the time interval in which Starburst99 ($Z=0.014$) is consistent with observations, its $M_{V,{\rm model}}$ varies from about $-13.2$ to $-13.4$. Likewise, $M_{V,{\rm model}}$ is approximately $-13.4$ for BPASS ($Z=0.014$) and $-13.5$ for BPASS ($Z=0.006$) when they are consistent with observations. Using the maximum and minimum values for $M_{V,{\rm model}}$ from all models ($-13.2$ and $-13.5$, respectively), we obtain an estimate of 4790--6310\,$M_{\odot}$ for the initial mass of a cluster consistent with the observed late-time source associated with SN~2010jp.  An important caveat, of course, is that at lower cluster masses, stochastic sampling of rare high-mass stars that have brief post-main-sequence phases may alter the colour for a given age. Since the age and initial mass from this cluster colour analysis agree with the age/mass estimate from assuming that the light of the remaining source is a single evolved star (Sec. 4.1), it seems plausible that the observed cluster colour for this range of cluster mass could be dominated by just one red supergiant (RSG).

We can compare the age inferred from the cluster colour to the lower limit for the age that comes from the assumption that all the light is provided by an individual supergiant star, as in Section \ref{sec:individualstar}.  Since a star of mass 15--20\,$M_{\odot}$ would give a luminosity consistent with the observed source at the location of SN~2010jp, a host cluster at this location could not contain any stars significantly more massive than this, or else the observed luminosity would be higher. A star of mass 20\,$M_{\odot}$ leaves the main sequence at 8.0\,Myr according to Geneva ($Z=0.014$) models, at 8.6\,Myr according to BPASS ($Z=0.014$) models, and at 8.8\,Myr according to BPASS ($Z=0.006$) models. We therefore place a lower bound of 8\,Myr on the age of the remaining late-time source at the location of SN 2010jp; the excluded range of ages is indicated by the blue shaded area at left in Figure \ref{fig:main}. This age is consistent with the range of ages derived by comparing the cluster colour to population-synthesis models, and the approximate age estimated from the H~{\sc ii} region radius.

\subsection{Likely age and initial mass of SN~2010jp} \label{sec:finalestimates}

In the case that the observed late-time source is an individual star born alongside the progenitor of SN~2010jp, the SN itself would have initially been only slightly more massive than the remaining supergiant star. Our estimate for the mass of an individual star consistent with the observed source was 15--20\,$M_{\odot}$ (Sec. \ref{sec:individualstar}). According to Geneva ($Z=0.014$) models, a single star of initial mass 20\,$M_{\odot}$ leaves the main sequence at 8.0\,Myr, whereas a 15\,$M_{\odot}$ star does so between 10.1 and 10.7\,Myr (the \cite{ekstrom2012} piecewise function used to estimate Geneva main-sequence lifetimes is discontinuous at 15\,$M_\odot$, so this number varies slightly depending on which formula is used). Assuming that $\sim 10$\% of the total stellar lifetime is spent post-main-sequence, these ages correspond to core-collapse times of 22.8\,$M_\odot$ and 15.9--17.1\,$M_\odot$ stars, respectively. Similarly, BPASS ($Z=0.014$) predicts main-sequence lifetimes of 8.6\,Myr for 20\,$M_\odot$ stars and 12.2\,Myr for 15\,$M_\odot$ stars, corresponding to core-collapse times of 22.1\,$M_\odot$ and 16.3\,$M_\odot$ stars, respectively. Finally, BPASS ($Z=0.006$) predicts main-sequence lifetimes of 8.8\,Myr for 20\,$M_\odot$ stars and 12.3\,Myr for 15\,$M_\odot$ stars, corresponding to times till core-collapse appropriate for stars of 22.0\,$M_\odot$ and 16.4\,$M_\odot$, respectively. Hence, we take 8\,Myr as a lower bound on the age of SN~2010jp's progenitor and 22\,$M_\odot$ as an upper bound on its mass in this case. We note that while the age range of 10--12\,Myr is appropriate for a single $\sim 15$\,$M_{\odot}$ star, the age could be older than this if the source has evolved through binary evolution or if the light arises from multiple stars.

In the more likely case that the late-time source at the location of SN~2010jp is indeed a star cluster (as argued in Sec.~\ref{sec:extendedemission}), we assume that the progenitor of SN~2010jp was coeval with the cluster, and so our estimate of the age of SN~2010jp from the cluster's colour is 8--13\,Myr.
As noted previously, the Starburst99 ($Z=0.014$) colour is consistent with observations from 8.4 to 12.7\,Myr. Again assuming 10\% of total stellar lifetime is spent post-main-sequence, this range corresponds to times till core-collapse for stars of 14.4--21.4\,$M_\odot$. Similarly, the BPASS ($Z=0.014$) colour is consistent with observations from 9.0 to 10.4\,Myr, corresponding to times till core-collapse for 18.6--21.1\,$M_\odot$ stars. Finally, the BPASS ($Z=0.006$) colour is consistent with observations from 9.7 to 11.5\,Myr, corresponding to times till core-collapse for 17.4--20.1\,$M_\odot$ stars. These results are in excellent agreement with the mass and age range inferred by assuming that the light is dominated by an individual evolved star. 
Thus, 14--22\,$M_{\odot}$ is the most probable range of initial masses indicated by the source's colour and luminosity.  

As noted earlier, the presence of an extended H~{\sc ii} region around the position of SN~2010jp may favour younger ages within this range.  Considering also the need for a late O-type star (at least $\sim 18$\,$M_{\odot}$) to produce the extended H~{\sc ii} region, we adopt 18--22\,$M_{\odot}$ as the most probable range of initial masses of the SN progenitor.    These quoted initial masses apply for single stars; if SN~2010jp evolved through binary evolution as a mass gainer or post-merger object, then it may have been more massive than 22\,$M_{\odot}$ at the time of explosion while still having the age of a single 22\,$M_{\odot}$ main-sequence star.  Relatively low-luminosity LBVs that are thought to be potential SN~IIn progenitors do reside in this mass range \citep{groh13,smith04,smith07,smithGaia19}.

This initial mass range around 20\,$M_{\odot}$ is interesting in the context of SN~2010jp's very unusual explosion properties.  As noted by \citet{smith2012}, the evidence for a jet-driven explosion, the relatively low peak luminosity (suggesting a low $^{56}$Ni production), and the remote location of SN~2010jp all seem reminiscent of expectations for some rare cases of weak SNe with fallback at low metallicity.  Note that \citet{smith2012} estimated an upper limit to the SN's $^{56}$Ni mass of 0.003\,$M_{\odot}$.   \citet{heger2003} expect such explosions over a wide range of subsolar metallicity for initial masses around 25\,$M_{\odot}$, while more recent studies find such events to be even more rare and pushed to somewhat higher masses ($\ga 27\,M_{\odot}$) in only a subset of models \citep{suk18}.  While the initial mass we infer for SN~2010jp is lower than this, that inferred initial mass is based on the age of a single star; the mass at the time of explosion may have behaved as a higher mass star if the SN~2010jp progenitor had gained mass through binary evolution.  An initial mass of $\la 20\,M_{\odot}$ may also be interesting because, as noted previously \citep{tim96,suk18,suk20}, this mass marks a transition from convective to radiative carbon burning that may produce divergent results in later burning phases and may strongly influence the fate of the star.  

Finally, an initial mass of $\sim 20\,M_{\odot}$ may have implications for the so-called RSG problem, if SN~2010jp was indeed the result of a peculiar faint explosion related to fallback with a jet (which might occur at the transition from neutron star to black hole remnants).  Namely, the inferred initial masses for detected RSG progenitors of nearby SNe~II-P (and several upper limits) have prompted suggestions that the most massive RSGs may be missing from the population of detected SNe because they collapse to a black hole without making a bright explosion.  The transition mass, if it exists, is still a matter of ongoing debate.  \citet{smartt09} estimate that RSGs with initial masses above 16.5\,$M_{\odot}$ collapse to a black hole without a bright SN explosion.  When circumstellar extinction or proper bolometric corrections are included, this number may shift up to around 19--21\,$M_{\odot}$ \citep{wal12,db18}.  Moreover, when one compares RSG luminosities to SN~II-P progenitor luminosities (rather than initial masses inferred through single-star model tracks), the steep luminosity function of RSGs indicates that the upper cutoff of RSG progenitors is not statistically significant \citep{db20}, so the question of such a transition mass remains open.  In any case, the initial mass for a very peculiar SN like SN~2010jp, if it arises from fallback or some other peculiar explosion, may directly impact this problem.

\section{Conclusions}

Late-time {\it HST} images at the location of the peculiar jet-driven SN~2010jp were analysed via photometry and PSF fitting to characterise the late-time source and constrain the mass and age of SN~2010jp itself. The possibility that the late-time source is due largely to ongoing CSM interaction from the SN event is dismissed owing to a lack of an observed decrease in continuum flux from 2012 to 2015, and because the H$\alpha$ emission appears to be extended. The possibility that the late-time source is an individual evolved star is considered, in which case this star would have had an initial mass of 15--20\,$M_{\odot}$. SN~2010jp can then be constrained to have an initial mass less than $22\,M_{\odot}$ and an age of at least 8\,Myr in this case. 

We suggest, however, that the late-time source is more likely the combined light from a star cluster given that H$\alpha$ emission is observed to be above the continuum, as well as evidence for an extended H~{\sc ii} region. In this case, comparing the observed $V-I$ colour with cluster evolution models allows us to estimate an age of 8--13\,Myr and a total mass of 4790--6310\,$M_{\odot}$ for the cluster. Lastly, the detection of an extended H~{\sc ii} regions suggests the presence of late O-type stars.  Altogether, we then estimate the progenitor of SN~2010jp to have had an initial mass 18--22\,$M_\odot$. As SN~2010jp is the first observed jet-driven SN~IIn, there is considerable interest in observing more SNe like it. If this species of SNe is characteristic of a particular type of environment and mass range, these results suggest that they may prefer relatively isolated, low mass, and moderately young star clusters.

\section*{Acknowledgements}

We thank an anonymous referee for helpful suggestions. Based on observations made with the NASA/ESA {\it Hubble SpaceTelescope}, obtained at the Space Telescope Science Institute (STScI), which is operated
by the Association of Universities for Research in Astronomy (AURA),
Inc., under NASA contract NAS 5-26555. Support was provided by
NASA through grants GO-13029 and GO-13787 from STScI. A.V.F. is also grateful for financial support from the TABASGO Foundation, the Miller Institute for Basic Research in Science (U.C. Berkeley, where he is a Senior Miller Fellow), the Christopher R. Redlich Fund, and many individual donors.  Some of the data presented in this paper were obtained from the Mikulski Archive for Space Telescopes (MAST).  This paper includes data gathered with the 6.5\,m Magellan Telescopes located at Las Campanas Observatory, Chile. 

Figures were created with Anaconda (Anaconda Software Distribution. Computer software. Vers. 2-2.4.0. Anaconda, Nov. 2016. Web. https://anaconda.com) using the following Python packages: Astropy (Astropy Collaboration et al. 2013; Price-Whelan et al. 2018 doi: 10.3847/1538-
3881/aabc4f); matplotlib Hunter (2007),DOI: 10.1109/MCSE.2007.55;  numpy van der Walt et al.  (2011), DOI : 10.1109/MCSE.2011.37.

Facilities: HST (WFC3), Magellan (IMACS).

\section*{Data Availability}

The data underlying this article will be shared on reasonable request
to the corresponding author.  {\it HST} data are nonproprietary and available from the public archive.




\bibliographystyle{mnras}
\bibliography{main} 









\bsp	
\label{lastpage}
\end{document}